\documentclass[english,journal]{IEEEtran}
\usepackage[T1]{fontenc}
\usepackage{color}
\usepackage{babel}
\usepackage{array}
\usepackage{multirow}
\usepackage{amsmath}
\usepackage{amssymb}
\usepackage{graphicx}
\usepackage[unicode=true,
 bookmarks=true,bookmarksnumbered=true,bookmarksopen=true,bookmarksopenlevel=1,
 breaklinks=false,pdfborder={0 0 0},backref=false,colorlinks=false]
 {hyperref}
\hypersetup{pdftitle={Your Title},
 pdfauthor={Your Name},
 pdfpagelayout=OneColumn, pdfnewwindow=true, pdfstartview=XYZ, plainpages=false}

\makeatletter

\providecommand{\tabularnewline}{\\}

\usepackage[caption=false,font=footnotesize]{subfig}

\newtheorem{assumption}{Assumption}
\newtheorem{theorem}{Theorem}
\newtheorem{remark}{Remark}
\newtheorem{definition}{Definition}
\newtheorem{ialgo}{Identification algorithm}
\newtheorem{calgo}{Optimization algorithm}

\newcommand{\bdefi}{\vspace{0.3em}\begin{definition}}
\newcommand{\edefi}{\end{definition}\vspace{0.3em}}
\newcommand{\bassu}{\vspace{0.3em}\begin{assumption}}
\newcommand{\eassu}{\end{assumption}\vspace{0.3em}}
\newcommand{\bthm}{\vspace{0.3em}\begin{theorem}}
\newcommand{\ethm}{\end{theorem}\vspace{0.3em}}
\newcommand{\brem}{\vspace{0.3em}\begin{remark}}
\newcommand{\erem}{\end{remark}\vspace{0.3em}}

\newcommand{\bialgo}{\vspace{0.3em} \noindent \rule{\columnwidth}{0.5pt} \begin{ialgo}\vspace{0.3em}}
\newcommand{\eialgo}{\end{ialgo} \rule{\columnwidth}{0.5pt}\vspace{0.3em}}

\newcommand{\bcalgo}{\vspace{0.3em} \noindent \rule{\columnwidth}{0.5pt} \begin{calgo}\vspace{0.3em}}
\newcommand{\ecalgo}{\end{calgo} \rule{\columnwidth}{0.5pt}\vspace{0.3em}}

\makeatother

\begin{document}

\title{Polynomial model inversion control:\\
numerical tests and applications}

\author{C. Novara\thanks{Carlo Novara is with Dipartimento di Automatica e Informatica, Politecnico
di Torino, Italy, e-mail: \protect\href{http://carlo.novara@polito.it}{carlo.novara@polito.it}.}}
\maketitle
\begin{abstract}
A novel control design approach for general nonlinear systems is described
in this paper. The approach is based on the identification of a polynomial
model of the system to control and on the on-line inversion of this
model. Extensive simulations are carried out to test the numerical efficiency of the approach. Numerical examples of applicative interest are presented, concerned
with control of the Duffing oscillator, control of a robot manipulator
and insulin regulation in a type 1 diabetic patient.
\end{abstract}

\section{Introduction}

Consider a nonlinear discrete-time system in regression form:
\begin{equation}
\begin{array}[t]{l}
y_{t}=h\left(u_{t}^{-},y_{t}^{-},\xi_{t}^{-}\right)\\
\\
u_{t}^{-}\doteq\left(u_{t-1},\ldots,u_{t-n}\right)\\
y_{t}^{-}\doteq\left(y_{t-1},\ldots,y_{t-n}\right)\\
\xi_{t}^{-}\doteq\left(\xi_{t-1},\ldots,\xi_{t-n}\right)
\end{array}\label{ss_sys}
\end{equation}
where $u_{t}\in U\subset\mathbb{R}^{n_{u}}$ is the known input, $y_{t}\in\mathbb{R}^{n_{y}}$
is the measured output, $\xi_{t}\in\Xi\subset\mathbb{R}^{n_{\xi}}$
is an unmeasured disturbance; $n$ is the system order; $U$ and $\Xi\doteq\left\{ \xi\in\mathbb{R}^{n_{\xi}}:\left\Vert \xi\right\Vert \leq\bar{\xi}\right\} $
are compact sets; the function $h$ is Lipschitz continuous on $\Omega_{h}\doteq Y^{n}\times U^{n}\times\Xi^{n}$,
where $Y$ is a compact set. $U$ accounts for possible constraints
on $u_{t}$.

Suppose that the system (\ref{ss_sys}) is unknown, but a set of noise-corrupted
measurements is available:
\begin{equation}
\mathcal{D}\doteq\left\{ \tilde{y}_{t},\tilde{u}_{t}\right\} _{t=1-L}^{0}\label{eq:data}
\end{equation}
where the tilde is used to denote the samples of the data set $\mathcal{D}$.

Let $\mathcal{Y}^{0}\subseteq Y^{n}$ be a set of initial conditions
of interest, $\mathcal{R}\doteq\left\{ \boldsymbol{r}=(r_{1},r_{2},\ldots):r_{t}\in Y,\forall t\right\} $
a set of output sequences of interest, and $\varXi\doteq\left\{ \boldsymbol{\xi}=(\xi_{1},\xi_{2},\ldots):\xi_{t}\in\Xi,\forall t\right\} $
the set of all possible disturbance sequences.\medskip{}

\emph{The problem is to design a controller for the system (\ref{ss_sys})
such that, for any $\boldsymbol{\xi}=(\xi_{1},\xi_{2},\ldots)\in\varXi$,
and for any initial condition $y_{0}\in\mathcal{Y}^{0}$, the output
sequence $\boldsymbol{y}=(y_{1},y_{2},\ldots)$ of the controlled
system tracks any reference sequence $\boldsymbol{r}=(r_{1},r_{2},\ldots)\in\mathcal{R}$.}

\medskip{}

To solve this problem, a novel data-driven control approach will be
described in the following, based on the identification of a polynomial
prediction model and on the online inversion of this model via the
efficient solution of suitable optimization problems. A simplified
version of the approach is presented in \cite{FoNo15}.

\section{Data-based prediction model}

\label{sec:ibc_approach}

A model is considered, of the form
\begin{equation}
y^{+}=f\left(u^{+},q^{-}\right)\label{eq:model}
\end{equation}
where $y^{+}\equiv y_{t}^{+}$ is a prediction of the system output
(over some finite time horizon), $u^{+}\equiv u_{t}^{+}$ is a vector
with the present and future input values and $q^{-}\equiv q_{t}^{-}\doteq\left(u_{t}^{-},y_{t}^{-}\right)$.
The subscript indicating the time will be omitted in the reminder
of the paper when not necessary. A parametric structure is taken for
the vector-valued function $f$. In particular, each component $f_{j}$
of $f$ is parametrized as
\begin{equation}
f_{j}\left(\cdot\right)=\sum_{i=1}^{N}\alpha_{ij}\phi_{i}\left(\cdot\right)\label{eq:bfe}
\end{equation}
where $\phi_{i}$ are polynomial basis functions, $\alpha_{ij}$ are
parameters to be identified and $j=1,\ldots,\tau n_{y}$. The parameters
$\alpha_{ij}$ can be identified from the data \eqref{eq:data} by
means of convex optimization.

\section{Polynomial inversion control}

\label{sec:c_alg}

The proposed control approach is based on the on-line inversion of
the model \eqref{eq:model}: at each time $t>0$, given a reference
sequence $r^{+}$ and the current regressor $q^{-}$, a command sequence
$u^{+}$ is looked for, such that the model output $\hat{y}^{+}$
is ``close'' to $r^{+}$: 
\begin{equation}
\hat{y}^{+}=f\left(u^{+},q^{-}\right)\cong r^{+}.\label{eq:inv1}
\end{equation}

Such a command sequence is found solving the optimization problem
\begin{equation}
u^{*}=\arg\min_{\mathfrak{u}\in U^{\tau}}J\left(\mathfrak{u},r^{+},q^{-}\right)\label{eq:opt2}
\end{equation}

where
\begin{equation}
J\left(\mathfrak{u},r^{+},q^{-}\right)\doteq\left\Vert r^{+}-f\left(\mathfrak{u},q^{-}\right)\right\Vert _{2}^{2}+\mu\left\Vert \mathfrak{u}\right\Vert _{2}^{2}\label{eq:objf2}
\end{equation}
and $\mu\geq0$ is a design parameter, determining the trade-off between
tracking precision and command activity.

The problem \eqref{eq:opt2} is solved at each sampling time, resulting
in the following control law:
\begin{equation}
u_{t}^{*}=u_{end}^{*}\equiv u_{end}^{*}\left(r_{t}^{+},q_{t}^{-}\right)\label{eq:ctr_law}
\end{equation}
where $u_{end}^{*}$ is the first entry of the vector $u^{*}$ in
\eqref{eq:opt2}.

The objective function \eqref{eq:objf2} is in general non-convex.
Moreover, the optimization problem \eqref{eq:opt2} has to be solved
on-line, and this may require a long time compared to the sampling
time used in the application of interest. To overcome these relevant
problems, three algorithms have been developed, allowing an efficient
computation of the optimal command input $u_{t}^{*}$ for the following
cases: 
\begin{enumerate}
\item SIMO system and piecewise constant command input; the optimal solution
can be computed ``almost analytically''.
\item MIMO system affine in $u^{+}$; the cost function is convex, implying
that the optimal solution can be obtained with ``low'' computational
cost.
\item General MIMO system. we will show below by means of extensive simulations
that the algorithm is able to find always a solution ``very close''
to a global one, in very short times. 
\end{enumerate}
The algorithms are based on a coordinate minimization approach but
are not described here.

\section{Optimization algorithm performance evaluation}

\label{sub:cca}

The optimization problem \ref{eq:opt2} was considered, where $f\left(\cdot\right)$
is a polynomial function of degree $d_{p}$ and $\mathfrak{u}\in U\subset\mathbb{R}^{m},\:r^{+},q^{-}\in\mathbb{R}^{m}$.
This problem is analogous to \eqref{eq:opt2} but the dependence on
time is not evidenced. The value $\mu=0$ was taken since, with this
value, if $r^{+}$ is in the range of $f\left(\cdot\right)$, we know
the global minimum of $J\left(\mathfrak{u},r^{+},q^{-}\right)$ to
be $0$.

Values of $m$ in the set $\{1,2,4,6,8\}$ and values of $d_{p}$
in the set $\{1,2,4,6\}$ were considered, corresponding to MIMO systems
with up to 8 command inputs and models with polynomial degree up to
6. Note that in all the applications presented below, degrees $2\div4$
led to a very satisfactory prediction and control performance. Degrees
larger than $4$ seem in general to not give any advantage.

For each combination of $m$ and $d_{p}$ in these sets, a Monte Carlo
simulation was carried out, consisting of 50 main trials, each consisting
of 100 sub trials (total number of trials: $5*4*50*100=100\,000$). 

In each main trial, $f\left(\cdot\right)$ was defined as a polynomial
function of degree $d_{p}$ with sparse random coefficients. In particular,
a number $n_{s}$ of nonzero coefficients was assumed, with $n_{s}$
ranging in the interval $[0,500]$ in function of $m$ and $d_{p}$
(the nonzero coefficients were chosen according to a Gaussian distribution
with zero mean and unitary variance). In each sub trial, a sequence
$r_{i}=f\left(u_{i}^{true},q_{i}^{-}\right)$ was generated, where
$q_{i}^{-}$ and $u_{i}^{true}$ are vectors with random entries (chosen
according to a uniform distribution with support $[-1,1]$), and $i=1,\ldots,100$.
Then, for each $i$, the optimization problem \eqref{eq:opt2} was
solved. Note that the decision variable $\mathfrak{u}$ is different
from the ``true'' input $u_{i}^{true}$. 

For each combination of the dimension $m$ and the polynomial degree
$d_{p}$, the following indexes were considered to evaluate the algorithm
performance:
\begin{itemize}
\item $E_{2}\doteq\frac{1}{5000}\sum_{i=1}^{5000}\left(J\left(u_{i}^{*},r_{i}^{+},q_{i}^{-}\right)-J\left(u_{i}^{true},r_{i}^{+},q_{i}^{-}\right)\right)$,
where $u_{i}^{*}$ is the solution of the optimization problem \eqref{eq:opt2},
computed for each random sample. Note that, in the present case, we
know that $J\left(u_{i}^{true},r_{i}^{+},q_{i}^{-}\right)=0$.
\item $E_{\infty}\doteq\underset{i=1,\ldots,5000}{\max}\left(J\left(u_{i}^{*},r_{i}^{+},q_{i}^{-}\right)-J\left(u_{i}^{true},r_{i}^{+},q_{i}^{-}\right)\right)$.
\item $T_{sc}\doteq$average time taken by a Matlab .m function to solve
a single optimization problem on a laptop with an i7 3Ghz processor
and 16 MB RAM. The average was computed over the $5000$ samples of
the Monte Carlo simulation. 
\item $T_{m}\doteq$average time taken by a compiled Simulink mex function
to solve a single optimization problem on the same laptop. This function
was generated in $10$ of the $50$ main trials, since this operation
is relatively complex. The average was thus computed over $10*100=1000$
samples of the Monte Carlo simulation.
\end{itemize}
The obtained results are summarized in Table \ref{tab:perf}. It can
be concluded that the coordinate descent minimization approach is
able to find precise solutions (i.e., giving small values of the objective
function) in short times for all the considered input dimensions and
polynomial degrees. It can also be observed that using compiled mex
functions allows a significant reduction of the computation times
for problems involving polynomials with a not too high degree in $\mathfrak{u}$.
A possible interpretation is that the Simulink automatic compiler
looses efficiency for large degree polynomials.

\begin{table}[tbh]
\centering%
\begin{tabular}{|c|c|c|c|c|c|c|}
\hline 
{\tiny{}$m$} & {\tiny{}$d_{p}$} & {\tiny{}$n_{s}$} & {\tiny{}$E_{2}$} & {\tiny{}$E_{\infty}$} & {\tiny{}$T_{sc}$ {[}s{]}} & {\tiny{}$T_{m}$ {[}s{]}}\tabularnewline
\hline 
\hline 
\multirow{4}{*}{{\tiny{}1}} & {\tiny{}1} & {\tiny{}3} & {\tiny{}1.2e-14} & {\tiny{}3.2e-14} & {\tiny{}2.7e-4} & {\tiny{}<1.0e-4}\tabularnewline
\cline{2-7} 
 & {\tiny{}2} & {\tiny{}6} & {\tiny{}1.9e-13} & {\tiny{}1.9e-12} & {\tiny{}3.0e-4} & {\tiny{}<1.0e-4}\tabularnewline
\cline{2-7} 
 & {\tiny{}4} & {\tiny{}15} & {\tiny{}2.1e-13} & {\tiny{}1.4e-12} & {\tiny{}3.4e-4} & {\tiny{}<1.0e-4}\tabularnewline
\cline{2-7} 
 & {\tiny{}6} & {\tiny{}28} & {\tiny{}1.1e-13} & {\tiny{}6.5e-13} & {\tiny{}3.7e-4} & {\tiny{}<1.0e-4}\tabularnewline
\hline 
\multirow{4}{*}{{\tiny{}2}} & {\tiny{}1} & {\tiny{}5} & {\tiny{}4.3e-12} & {\tiny{}1.6e-11} & {\tiny{}8.7e-4} & {\tiny{}<1.0e-4}\tabularnewline
\cline{2-7} 
 & {\tiny{}2} & {\tiny{}15} & {\tiny{}5.0e-3} & {\tiny{}0.048} & {\tiny{}1.6e-3} & {\tiny{}1.4e-4}\tabularnewline
\cline{2-7} 
 & {\tiny{}4} & {\tiny{}45} & {\tiny{}4.1e-3} & {\tiny{}0.022} & {\tiny{}2.2e-3} & {\tiny{}5.6e-4}\tabularnewline
\cline{2-7} 
 & {\tiny{}6} & {\tiny{}81} & {\tiny{}5.2e-3} & {\tiny{}0.034} & {\tiny{}5.4e-3} & {\tiny{}>$T_{sc}$}\tabularnewline
\hline 
\multirow{4}{*}{{\tiny{}4}} & {\tiny{}1} & {\tiny{}9} & {\tiny{}7.5e-5} & {\tiny{}4.2e-4} & {\tiny{}7.8e-4} & {\tiny{}<1.0e-4}\tabularnewline
\cline{2-7} 
 & {\tiny{}2} & {\tiny{}45} & {\tiny{}8.2e-3} & {\tiny{}0.039} & {\tiny{}3.1e-3} & {\tiny{}4.5e-4}\tabularnewline
\cline{2-7} 
 & {\tiny{}4} & {\tiny{}116} & {\tiny{}0.013} & {\tiny{}0.047} & {\tiny{}0.038} & {\tiny{}>$T_{sc}$}\tabularnewline
\cline{2-7} 
 & {\tiny{}6} & {\tiny{}197} & {\tiny{}0.014} & {\tiny{}0.046} & {\tiny{}0.17} & {\tiny{}>$T_{sc}$}\tabularnewline
\hline 
\multirow{4}{*}{{\tiny{}6}} & {\tiny{}1} & {\tiny{}13} & {\tiny{}4.3e-4} & {\tiny{}9.7e-4} & {\tiny{}1.9e-3} & {\tiny{}<1.0e-4}\tabularnewline
\cline{2-7} 
 & {\tiny{}2} & {\tiny{}81} & {\tiny{}0.013} & {\tiny{}0.048} & {\tiny{}0.011} & {\tiny{}1.2e-3}\tabularnewline
\cline{2-7} 
 & {\tiny{}4} & {\tiny{}197} & {\tiny{}0.016} & {\tiny{}0.048} & {\tiny{}0.21} & {\tiny{}>$T_{sc}$}\tabularnewline
\cline{2-7} 
 & {\tiny{}6} & {\tiny{}339} & {\tiny{}0.021} & {\tiny{}0.049} & {\tiny{}0.76} & {\tiny{}>$T_{sc}$}\tabularnewline
\hline 
\multirow{4}{*}{{\tiny{}8}} & {\tiny{}1} & {\tiny{}17} & {\tiny{}5.0e-4} & {\tiny{}8.6e-4} & {\tiny{}2.4e-3} & {\tiny{}<1.0e-4}\tabularnewline
\cline{2-7} 
 & {\tiny{}2} & {\tiny{}116} & {\tiny{}0.019} & {\tiny{}0.048} & {\tiny{}0.10} & {\tiny{}3.1e-3}\tabularnewline
\cline{2-7} 
 & {\tiny{}4} & {\tiny{}289} & {\tiny{}0.027} & {\tiny{}0.049} & {\tiny{}1.6} & {\tiny{}>$T_{sc}$}\tabularnewline
\cline{2-7} 
 & {\tiny{}6} & {\tiny{}500} & {\tiny{}0.032} & {\tiny{}0.049} & {\tiny{}8.3} & {\tiny{}>$T_{sc}$}\tabularnewline
\hline 
\end{tabular}\medskip{}

\protect\caption{Monte Carlo simulation results.}
\label{tab:perf}
\end{table}

\section{Applications}

\subsection{Duffing oscillator}

The Duffing system is a second-order damped oscillator with nonlinear
spring, described by the following differential equations:
\begin{equation}
\begin{array}{l}
\dot{x}_{1}=x_{2}\\
\dot{x}_{2}=-\alpha_{1}x_{1}-\alpha_{2}x_{1}^{3}-\beta x_{2}+u\\
y=x_{1}+\xi
\end{array}\label{duffing}
\end{equation}
where $x=(x_{1},x_{2})$ is the system state ($x_{1}$ and $x_{2}$
are the oscillator position and velocity, respectively), $u$ is the
input, $y$ is the output, and $\xi$ is a zero-mean Gaussian noise
having a noise-to-signal standard deviation ratio of $0.03$. The
following values of the parameters have been considered: $\alpha_{1}=-1$,
$\alpha_{2}=1$, $\beta=0.2$. For these parameter values and for
certain choices of the input signal, this system exhibits a chaotic
behavior, and this makes control design a particularly challenging
problem.

A simulation of the Duffing system (\ref{duffing}) having duration
$400$ s was performed, using the input signal $u(\tau)=0.3\sin(\tau)+\xi^{u}(\tau)$,
where$\tau$ here denotes the continuous time and $\xi^{u}(\tau)$
is a white Gaussian noise with zero mean and standard deviation $0.2$.
\medskip{}

A set of $L=4000$ data were collected from this simulation with a
sampling period $T_{s}=0.1$ s:
\[
\mathcal{D}\doteq\left\{ \tilde{u}_{t},\tilde{y}_{t}\right\} _{t=-1999}^{0}
\]
where $\widetilde{u}_{t}=u(T_{s}t)$ are the measurements of the input
and $\widetilde{y}_{t}=y(T_{s}t)$ are the measurements of the output.\medskip{}

A nonlinear controller was designed following the approach described
in Sections \ref{sec:ibc_approach} and \ref{sec:c_alg}. This controller
was applied to the Duffing system (\ref{duffing}). \medskip{}

\begin{figure}[h]
\includegraphics[scale=0.6]{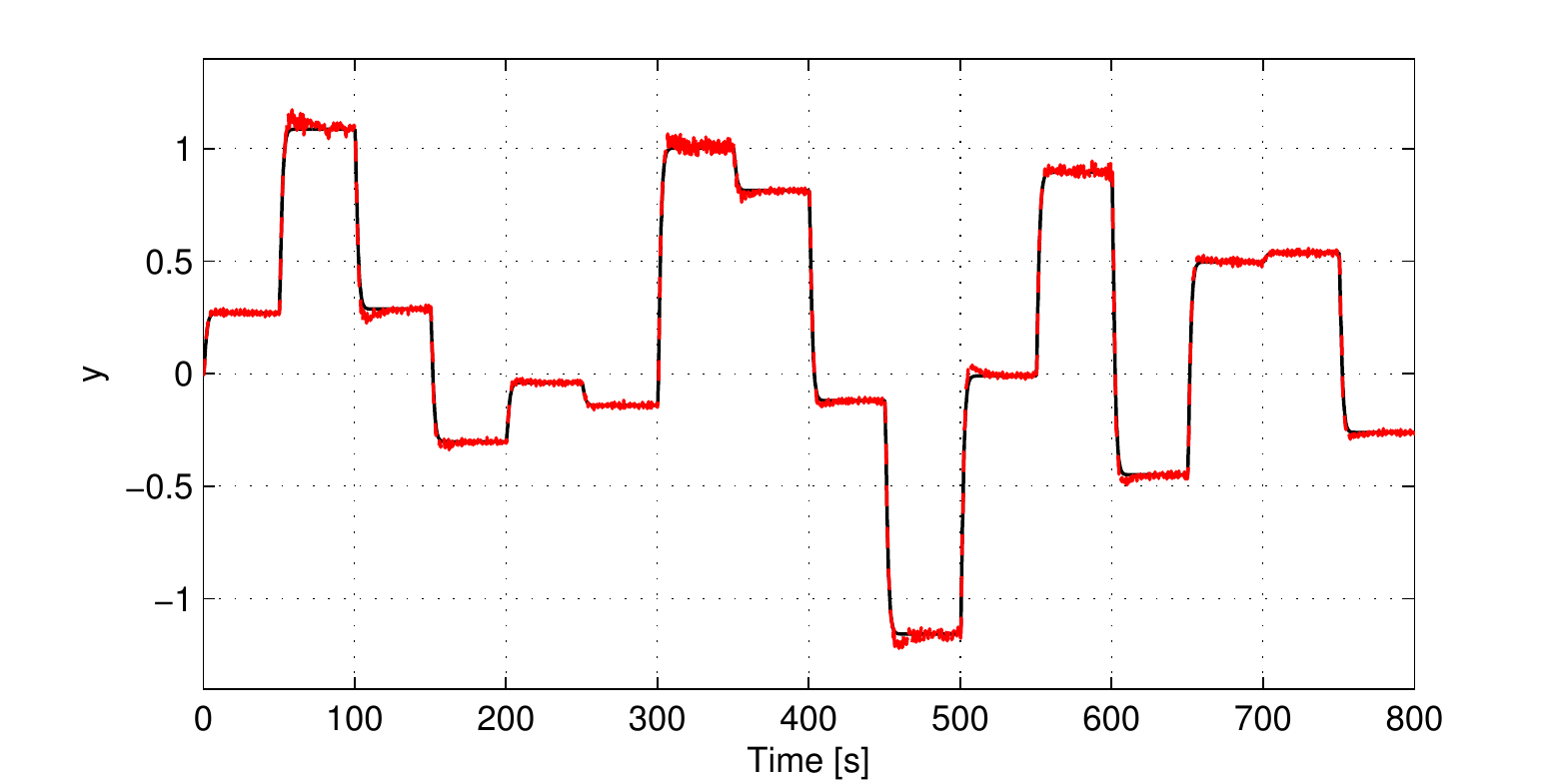}

\protect\caption{Tracking performance of the controlled system. Continuous (black)
line: reference. Dashed (red) line: actual output.}

\label{comp} 
\end{figure}

A testing simulation of the controlled system with duration $800$
s was performed, using zero initial conditions and a reference signal
$r_{t}$ generated as a sequence of random steps, filtered by a second-order
filter with a cutoff frequency of $2$ rad/s (this filter has been
inserted in order to ensure not too abrupt variations). A Gaussian
noise affecting the output measurements, having zero-mean and a noise-to-signal
standard deviation ratio of $0.03$ was included in the simulation.
In Figure \ref{comp}, the output of the controlled system is compared
to the reference. \medskip{}

Then, a Monte Carlo simulation was carried out, where this data-generation-control-design-and-testing
procedure was repeated 100 times. For each trial, the tracking performance
was evaluated by means of the Root Mean Square tracking error 
\[
RMS\doteq\sqrt{\frac{1}{8000}\sum\nolimits _{t=1}^{8000}\left(r_{t}-y_{t}\right)^{2}}.
\]
The average $RMS$ error obtained in the Monte Carlo simulation is
$\overline{RMS}=0.015$. 

\begin{figure}[h]
\includegraphics[scale=0.6]{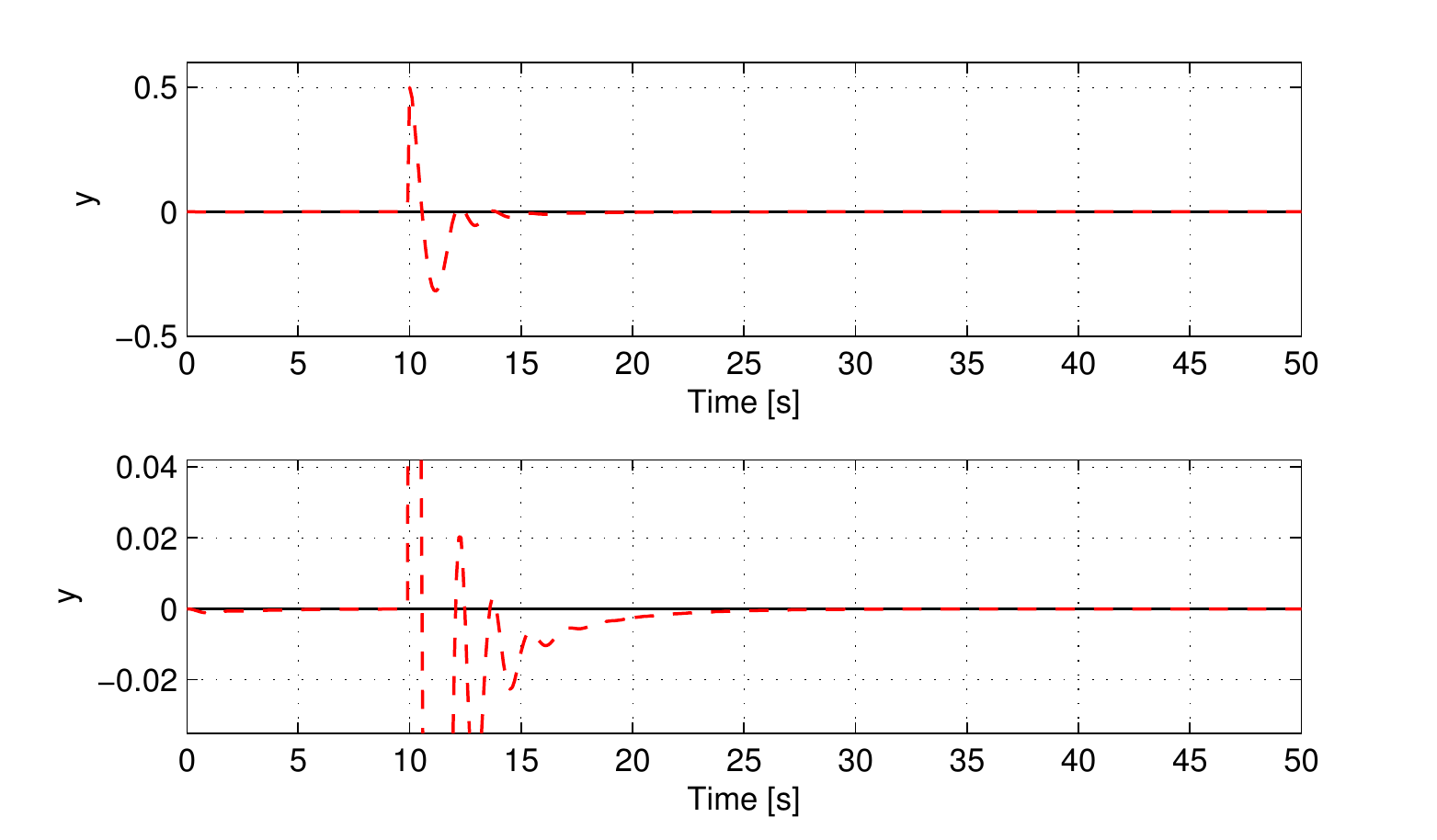}

\protect\caption{Above: disturbance rejection of the controlled system. Below: same
figure, with zoomed $y$ axis. Continuous (black) line: reference.
Dashed (red) line: actual output. }

\label{comp-2} 
\end{figure}

A simulation of the closed-loop system was also performed where $r_{t}=0,\:\forall t$
and $\xi_{t}$ was a step disturbance of amplitude $0.5$. The output
signals obtained in these simulations are shown in Figure \ref{comp-2}.\medskip{}

From these results, it can be concluded that the designed controller
is able to (1) ensure a very accurate tracking, even in the presence
of quite significant measurement noises; (2) reject/attenuate strong
step disturbances.

\subsection{Robot manipulator}

\begin{figure}[h]
\begin{centering}
\includegraphics[scale=0.7]{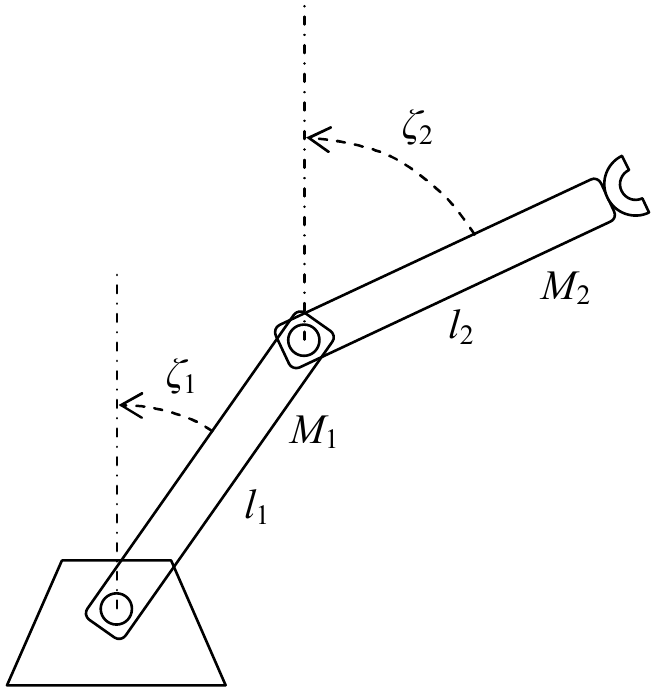} 
\par\end{centering}

\protect\caption{Robot Manipulator. }

\label{fig20} 
\end{figure}

The 2-DOF (2-degrees of freedom) robot manipulator depicted in Figure
\ref{fig20} has been considered, where $\zeta_{1}$ and $\zeta_{2}$
are the angular positions of the two segments of the robot arm, $u_{1}$
and $u_{2}$ are the control torques acting on these segments, $l_{1}$
and $l_{2}$ are the segment lengths, and $M_{1}$ and $M_{2}$ are
the segment masses. The parameter values $l_{1}=0.8$ m, $l_{2}=0.7$
m, $M_{1}=2.5$ Kg, $M_{2}=2$ Kg have been assumed.

This robot manipulator is a MIMO system (with 2 inputs and 2 outputs),
described by the following continuous-time state-space nonlinear equations:
\begin{equation}
\begin{array}{l}
\dot{z}(\tau)=A^{c}(z(\tau))z(\tau)+B^{c}(z(\tau))u(\tau)\\
y(\tau)=\left[\begin{array}{c}
z_{1}(\tau)\\
z_{2}(\tau)
\end{array}\right]
\end{array}\label{robot_ss}
\end{equation}
where $\tau$ is the continuous time, $z(\tau)=[\zeta_{1}(\tau)$
$\zeta_{2}(\tau)$ $\dot{\zeta}_{1}(\tau)$ $\dot{\zeta}_{2}]^{\top}$
is the state, $u(\tau)=[u_{1}(\tau)$ $u_{2}(\tau)]^{\top}$ is the
input, and the expressions of $A^{c}(z(\tau))\in\mathbb{R}^{4\times4}$
and $B^{c}(z(\tau))\in\mathbb{R}^{4\times2}$ can be found in \cite{KwWe05}.

A set of $L=5000$ data was generated by simulation of (\ref{robot_ss}):
\[
\mathcal{D}\doteq\left\{ \tilde{y}_{t},\tilde{u}_{t},\right\} _{k=-4999}^{0}.
\]
The data were collected with a sampling time $T_{s}=0.02$ s, using
the following input signals:
\begin{equation}
u_{j}(\tau)=\left\{ \begin{array}{l}
-20z_{j}(\tau),\text{ if }\left\vert z_{j}(\tau)\right\vert \geq1.75\text{ rad}\\
0,\text{\quad if }l<\tau\leq l+500,\text{ }l=500,1500,2500,3500,\\
\qquad\text{and }\left\vert z_{j}(\tau)\right\vert <1.75\\
U\sin(\omega_{j1}\tau)+U\sin(\omega_{j2}\tau),\text{\quad otherwise,}
\end{array}\right.\label{u_rob}
\end{equation}
where $j=1,2$, $U=\mathrm{rand}[50,150]$ Nm, $\omega_{11}=\mathrm{rand}[0.05,0.09]$
rad/s, $\omega_{12}=\mathrm{rand}[0.5,0.11]$ rad/s $\omega_{21}=\mathrm{rand}[0.04,0.1]$
rad/s $\omega_{11}=\mathrm{rand}[0.7,1.2]$ rad/s. The notation $U=\mathrm{rand}[50,150]$
means that $U$ is a number, randomly chosen according to a uniform
distribution in the interval $[50,150]$. The feedback input on the
first line of (\ref{u_rob}) was applied in order to limit the working
range of $z_{1}$ and $z_{2}$ to the interval $[-\pi,\pi]$ rad (the
gain $-20$ and the threshold $1.75$ rad were chosen thorough several
preliminary simulations). Measurement noises were added to $y_{j}$,
$j=1,2,$ simulated as uniform noises with amplitude $0.02$ rad.\medskip{}

From these data, two controllers were designed following the approach
described in Sections \ref{sec:ibc_approach} and \ref{sec:c_alg}:
The first one is based on a general nonlinear prediction model. The
second one is based on a prediction model affine in $u^{+}$. For
comparison, the controller in \cite{ChLPV11} has been considered,
designed by means of a two-step method, consisting in LPV model identification
and Gain Scheduling (GS) design.\medskip{}

\begin{figure}[h]
\begin{centering}
\includegraphics[scale=0.6]{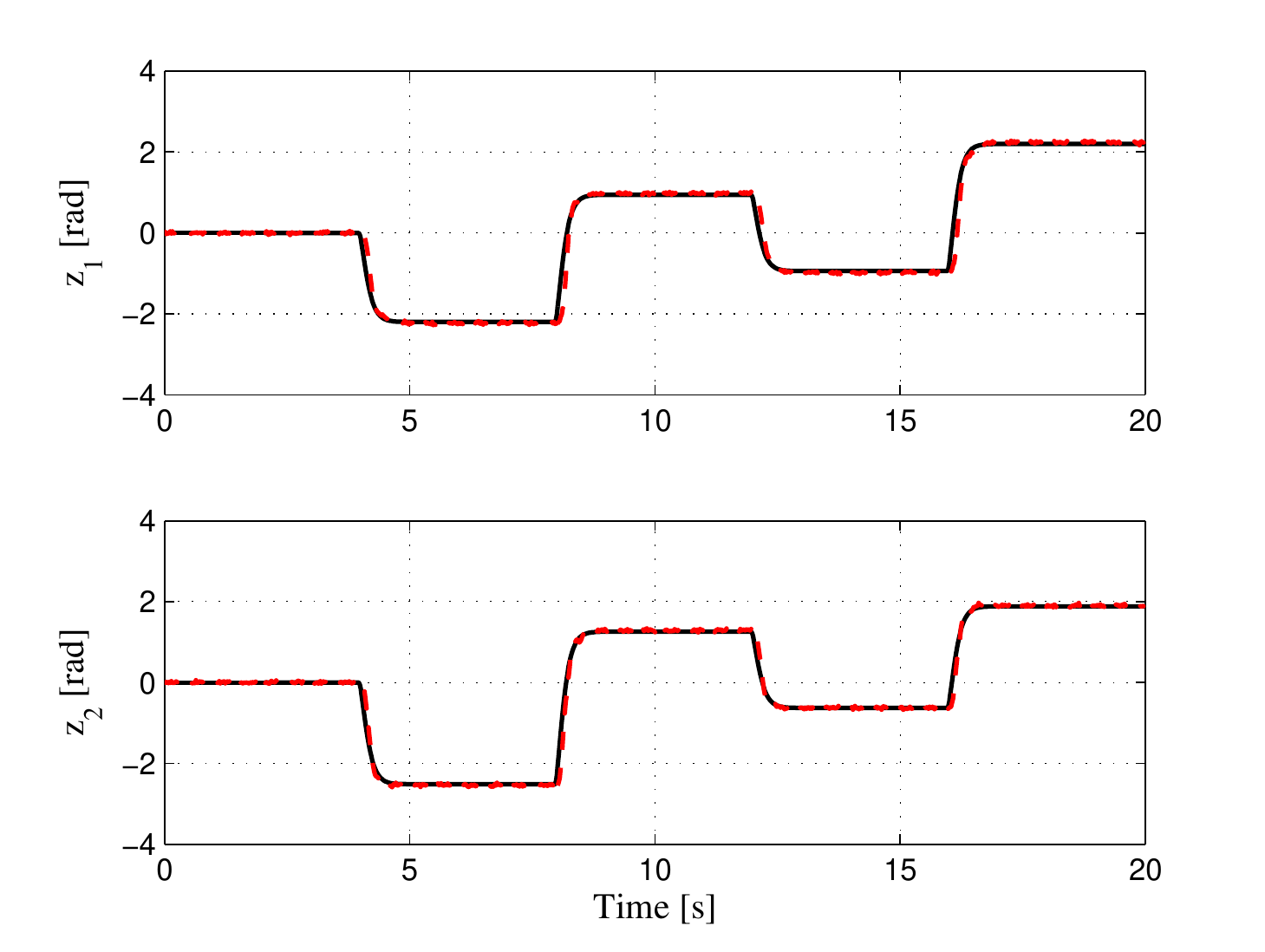} 
\par\end{centering}

\protect\caption{Robot Manipulator. Continuous (black) line: reference. Dashed (red)
line: closed-loop system output.}

\label{fig22} 
\end{figure}

\begin{figure}[h]
\begin{centering}
\includegraphics[scale=0.6]{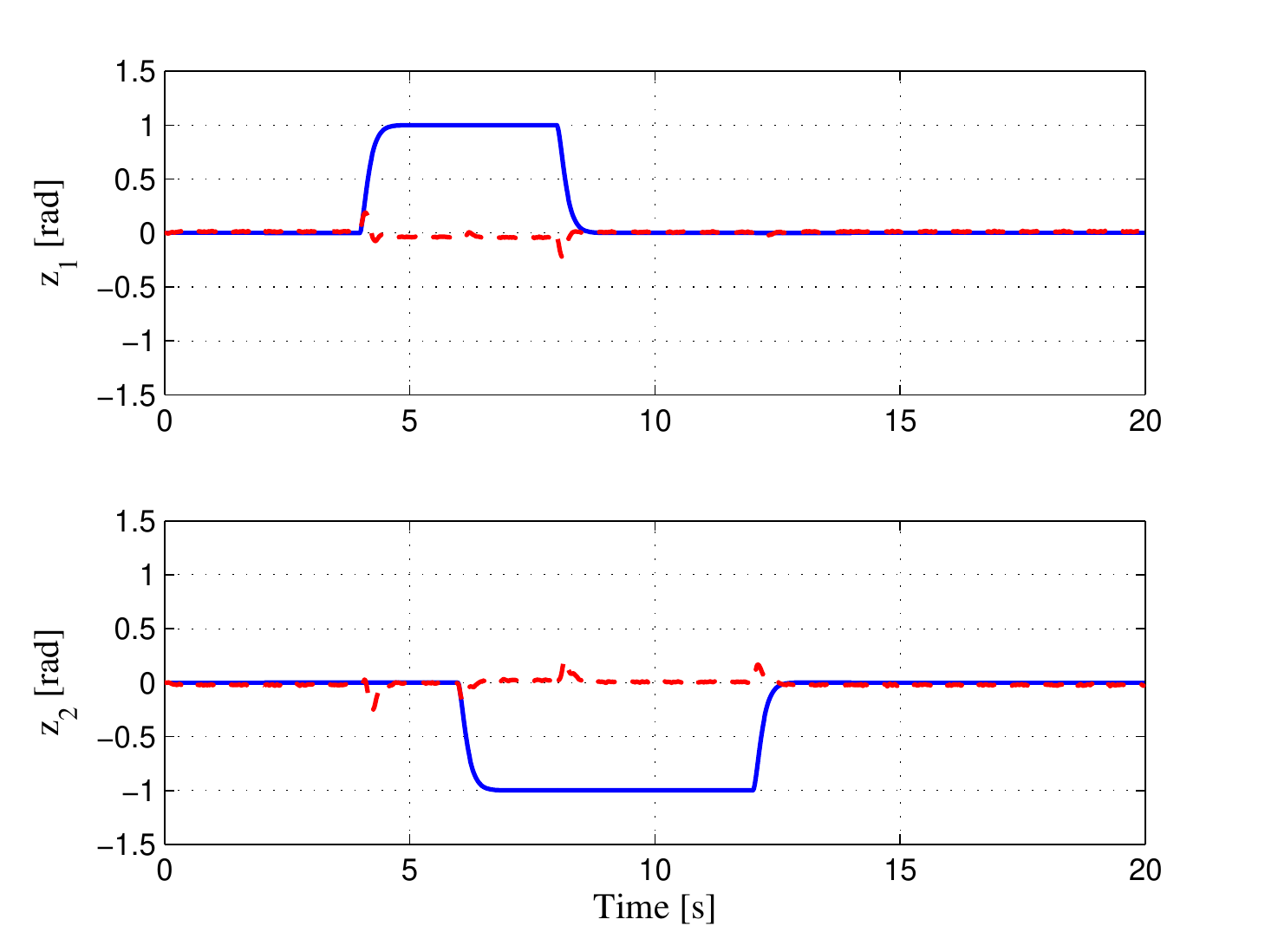} 
\par\end{centering}

\protect\caption{Robot Manipulator. Continuous (blue) line: disturbance. Dashed (red)
line: closed-loop system output.}

\label{fig22-1} 
\end{figure}

A first simulation was performed to test all the controllers in the
task of reference tracking. Zero initial conditions were assumed.
A reference signal of length $5000$ samples (corresponding to $100$
s) was used, defined as a random sequence of step signals with amplitudes
in the interval $[-\pi,\pi]$, filtered by a second-order filter with
a cutoff frequency of $10$ rad/s. This filter was inserted in order
to ensure not too high variations. The outputs were corrupted by random
uniform noises with amplitude $0.02$ rad. In Figure \ref{fig22},
the angular positions of the closed-loop system with the first controller
are compared with the references for the first $20$ s of this simulation.
Note that the two position references were chosen quite similar to
each other (but not equal) in order to allow the manipulator to reach
in a simple way any position in its range.\textcolor{red}{{} }A second
simulation was performed to test the controllers in the task of disturbance
attenuation.\textcolor{red}{{} }Zero initial conditions and a zero reference
were assumed. An output disturbance signal of length $1000$ samples
(corresponding to $20$ s) was considered, defined as a sequence of
two steps (one for each output channel) of amplitude $1$ rad, filtered
by a second-order filter with a cutoff frequency of $10$ rad/s. The
outputs were also corrupted by random uniform noises with amplitude
$0.02$ rad. In Figure \ref{fig22-1}, the angular positions of the
closed-loop system with the first controller are shown, together with
the disturbance signals.\medskip{}

Then, a Monte Carlo (MC) simulation was carried out, where this procedure
(data generation, control design, reference tracking test) was repeated
200 times. For each trial, the tracking performance was evaluated
by means of the Root Mean Square tracking errors, defined as
\[
RMS_{i}\doteq\sqrt{\frac{1}{5000}\sum\nolimits _{t=1}^{5000}\left(r_{i,t}-y_{i,t}\right)^{2}},\:i=1,2,
\]
where $r_{i,t}$ is the $i$th component of the reference signal and
$y_{i,t}$ is the $i$th component of the controlled system output.
The average errors $\overline{RMS}_{i}$ obtained in the MC simulation
are reported in Table \ref{tab1-1}. From these results, it can be
concluded that the designed control systems are quite effective, showing
a fast and precise tracking, and a significant disturbance attenuation
capability. In comparison with the two-step method of \cite{ChLPV11},
the proposed approach is simpler, since a polynomial model of the
form \eqref{eq:model} has in general a significantly simpler structure
wrt an LPV model (and, in particular, wrt a state-space LPV model).
Moreover, the tracking results obtained by the inversion-based controllers
are similar (or even slightly better) than those obtained by the GS
controller, despite the fact that this latter uses a stronger information
on the system \eqref{robot_ss} (i.e., the information that \eqref{robot_ss}
is a quasi-LPV system).

The computational times for the control design phase (referred to
a laptop with an i7 3Ghz processor and 16 MB RAM) resulted quite low,
considering that the set used for design consists of 5000 data: 92
s (nonlinear model), 83 s (affine model). The control algorithm on-line
evaluation times resulted also quite low: 2.1e-3 s (nonlinear model),
1.0e-3 s (affine model). This shows that these algorithms can be effectively
implemented on real time processors.

\begin{table}[tbh]
\centering

\begin{tabular}{|c||c|c|c|}
\hline 
 & controller 1 & controller 2 & GS\tabularnewline
\hline 
\hline 
$\overline{RMS}_{1}$ & 0.159 & 0.160 & 0.167\tabularnewline
\hline 
$\overline{RMS}_{2}$ & 0.114 & 0.115 & 0.152\tabularnewline
\hline 
\end{tabular}\smallskip{}

\protect\caption{Robot Manipulator. Average $RMS$ tracking errors.}

\label{tab1-1}
\end{table}

\subsection{Type 1 diabetes}

A model representing a type 1 diabetic patient has been considered
in this example. The inputs of this model are the carbohydrate-based
meal input and the insulin input function, the output is the blood
glucose concentration (glycemic response). The model state equations
are the following:

\begin{equation}
\begin{array}{ccl}
\frac{dy(t)}{dt} & = & -\left[p_{1}+\eta(t)\right]y(t)+p_{1}G_{b}+\frac{1}{V_{g}}w(t)\\
\frac{d\eta(t)}{dt} & = & -p_{2}\eta(t)+p_{3}[I(t)-I_{b}]\\
\frac{dI(t)}{dt} & = & \frac{k_{a}}{V_{d}}I_{2}(t)-k_{e}I(t)\\
\frac{dI_{1}(t)}{dt} & = & -k_{21}I_{1}(t)+\frac{1}{V_{I}}u(t)\\
\frac{dI_{2}(t)}{dt} & = & k_{21}I_{1}(t)-(k_{d}+k_{a})I_{2}(t)
\end{array}\label{eq:patient}
\end{equation}
where $y(t)$ is the blood glucose concentration (the system output),
$I(t)$ is the blood insulin concentration, $\eta(t)$ is the insulin
concentration in a remote compartment, $V_{g}$ is the volume distribution,
$w(t)$ is the carbohydrate-based meal input (the system unmeasured
input), $I_{1}(t)$ is the subcutaneous insulin mass in the injection
depot, $I_{2}(t)$ is the subcutaneous insulin mass proximal to plasma
and $u(t)$ is the injected insulin rate (the system measured input);
$p_{1}$, $p_{2}$, $p_{3}$ are individual subject parameters, $V_{d}$
is the plasma distribution volume, $k_{21}$, $k_{a}$, $k_{d}$,
and $k_{e}$ are insulin pharmacokinetic parameters, $I_{b}$ is the
basal blood insulin concentration and $G_{b}$ is the basal blood
glucose concentration. 

The first two equations of (\ref{eq:patient}), describing the glucose
dynamics, have been taken from the Bergman model, \cite{Bergman};
the last three equations of (\ref{eq:patient}), describing the insulin
kinetics, have been taken from the Shimoda model, \cite{Nucci}. The
following parameter values have been assumed: $p_{1}=0.031\:min^{-1}$,
$p_{2}=0.012\:min^{-1}$, $p_{3}=9.56e-6\:min^{-2}mL/\mu U$, $V_{g}=1.45\:dL/kg$,
$V_{d}=0.2\:mL/kg$, $V_{I}=5e-3\:mL$, $k_{21}=0.0166\:min^{-1}$,
$k_{a}=0.0133\:min^{-1}$, $k_{d}=0.0033\:min^{-1}$, $k_{e}=0.3\:min^{-1}$,
$I_{b}=0\:\mu U/mL$ and $G_{b}=180\:mg/dL$. In this simulated example,
the model (\ref{eq:patient}) represents the unknown ``true'' patient
metabolic system to control.\medskip{}

It must be remarked that the model (\ref{eq:patient}) is not the
most recent that can be found in the literature and may also be not
sufficiently adequate to describe a real diabetes patient. However,
the aim of this numerical example is to test the proposed control
algorithm on a non trivial nonlinear system and thus the particular
choice of the model used as the ``true'' system is not relevant. \medskip{}

A simulation of the patient system (\ref{eq:patient}) was performed,
where the insulin input was taken from a set of experimental data,
measured on a real patient. The meal input was simulated as a superposition
(with positive coefficients) of exponentially decaying signals $w_{j}(t)\:j=1,2,\ldots$,
where each contribution $w_{j}(t)$ represents a single meal. These
signals are of the form
\begin{equation}
w_{j}(t)=\left\{ \begin{array}{ll}
0, & t<t_{j}\\
\left(t-t_{j}\right)e^{-0.6\left(t-t_{j}\right)}, & t\geq t_{j}
\end{array}\right.\label{eq:dist}
\end{equation}
where $t_{j}$ is the time at which the patient started to eat. The
times $t_{j}$ were realistically chosen in order to have an insulin
injection a few minutes before a meal. A negative term of the form
\eqref{eq:dist} were also added to the meal input in order to reproduce
the effects of an external input yielding a decrease of the output
(e.g. a physical activity). The output signal (the blood glucose concentration)
resulting from this simulation was corrupted by a white noise, having
a noise-to-signal standard deviation ratio of $3\%$.

A set of $L=4800$ data (corresponding to 10 days) was collected from
this simulation, using a sampling time $T_{s}=3$ $min$: 
\[
\mathcal{D}\doteq\left\{ \tilde{u}_{t},\tilde{y}_{t}\right\} _{t=-4799}^{0}
\]
where $\widetilde{u}_{t}=u(T_{s}t)$ are the measurements of the insulin
input and $\widetilde{y}_{t}=y(T_{s}t)$ are the measurements of the
output. Note that, as it happens in most real situations, the meal
input was not measured.\medskip{}

A nonlinear controller was designed following the approach described
in Sections \ref{sec:ibc_approach} and \ref{sec:c_alg}. This controller
was applied to the diabetes system \eqref{eq:patient}. \medskip{}

Three simulations of the patient system (\ref{eq:patient}) with duration
10 days were performed, using a meal input signal different from that
used to generate the design data $\mathcal{D}$. The insulin signal
was generated as follows:
\begin{itemize}
\item first simulation: zero insulin;
\item second simulation: insulin injected by the patient on the basis of
his/hers experience;
\item third simulation: insulin signal computed by the designed controller.
\end{itemize}
In the simulations, the output signal was corrupted by a white noise,
with a noise-to-signal standard deviation ratio of $3\%$.

\begin{figure}[tbh]
\includegraphics[scale=0.45]{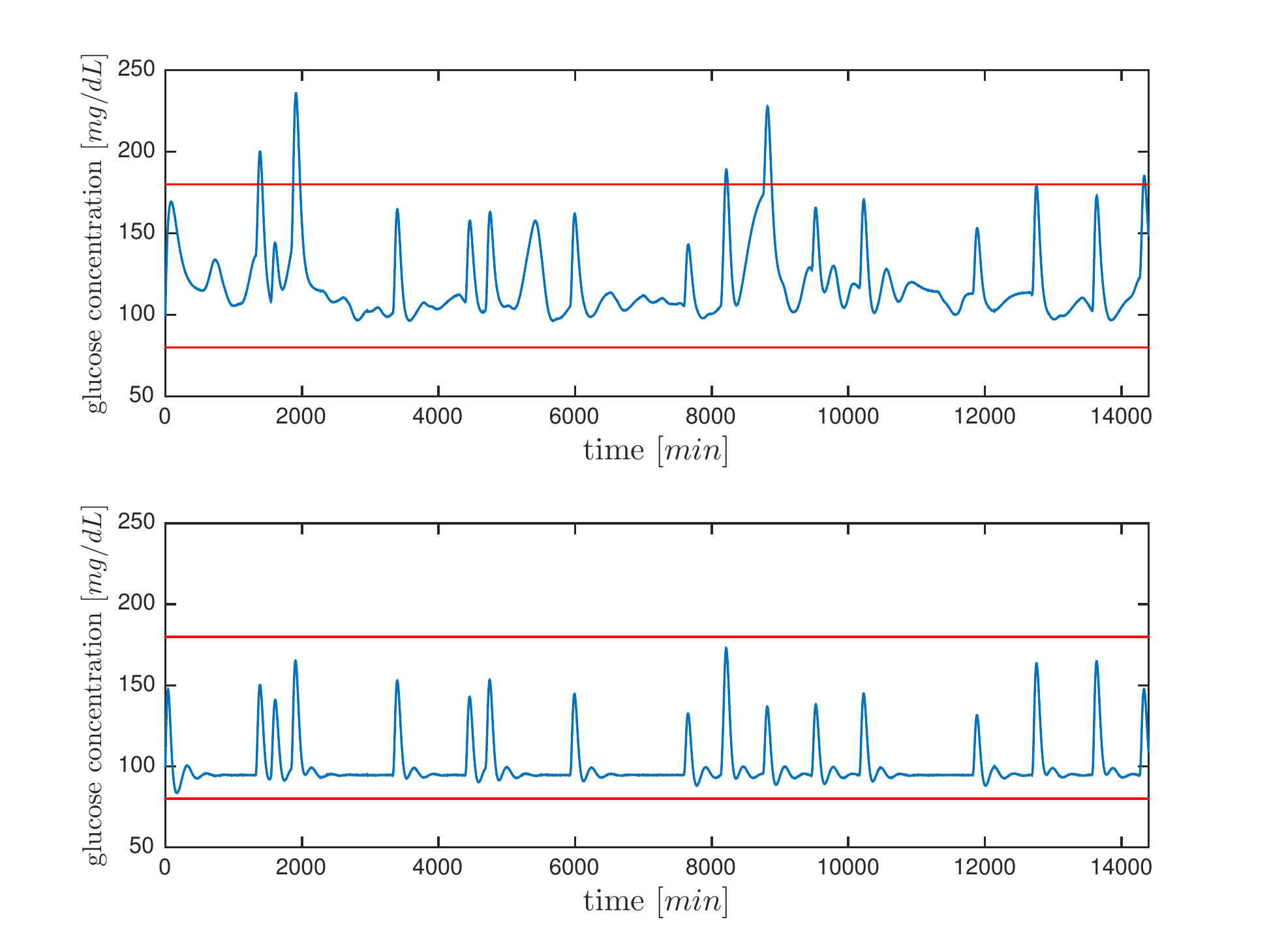}

\protect\caption{Blue lines: patient glucose concentration; red lines: safety bounds.
Above: result with insulin signal decided by the patient; below: result
with insulin signal generated by the controller. }

\label{fig_dab}
\end{figure}

The obtained results can be commented as follows: With no insulin,
the glucose concentration becomes very large, leading to serious health
problems of the patient. When the amount of injected insulin is decided
by the patient, the glucose concentration is somewhat regulated but
it may reach large values, which may worsen the patient health conditions
(see Figure \ref{fig_dab}). When the amount of injected insulin is
decided by the controller, the glucose concentration is always kept
within the interval $[80,180]$ \emph{$mg/dL$} which, in diabetes
treatment medicine, is commonly considered a safe interval (see Figure
\ref{fig_dab}).

\bibliographystyle{plain}
\bibliography{diabetes,lettnos_chapters,lpv,lettnos_journals}

\begin{thebibliography}{1}

\bibitem{Bergman}
R~N Bergman, L~S Phillips, and C~Cobelli.
\newblock Physiologic evaluation of factors controlling glucose tolerance in
  man: measurement of insulin sensitivity and beta-cell glucose sensitivity
  from the response to intravenous glucose.
\newblock {\em The Journal of Clinical Investigation}, 68(6):1456--1467, 12
  1981.

\bibitem{FoNo15}
S.~Formentin, C.~Novara, S.M. Savaresi, and M.~Milanese.
\newblock Active braking control system design: the d2-ibc approach.
\newblock {\em IEEE/ASME Transactions on Mechatronics}, 20(4):1573--1584, 2015.

\bibitem{KwWe05}
A.~Kwiatkowski and H.~Werner.
\newblock {LPV} control of a {2-DOF} robot using parameter reduction.
\newblock In {\em Proceedings of the {IEEE} Conference on Decision and Control
  and European Control Conference}, Seville, Spain, 2005.

\bibitem{ChLPV11}
C.~Novara.
\newblock Set membership identification of state-space {LPV} systems.
\newblock In P.~Lopes dos Santos, T.P.~Azevedo Perdico\'{u}lis, C.~Novara, J.A.
  Ramos, and D.E. Rivera, editors, {\em Linear Parameter-Varying System
  Identification -- New Developments and Trends, Advanced Series in Electrical
  and Computer Engineering Vol. 14}, pages 65--93. World Scientific, 2011.

\bibitem{Nucci}
Gianluca Nucci and Claudio Cobelli.
\newblock Models of subcutaneous insulin kinetics. a critical review.
\newblock {\em Computer Methods and Programs in Biomedicine}, 62(3):249 -- 257,
  2000.

\end{thebibliography}

\end{document}